\documentclass[12pt]{article}
\usepackage{chatsagnik}
\usepackage{framed}
\usepackage{xcolor}

\colorlet{shadecolor}{orange!15}
\renewenvironment{quote}{\begin{shaded*}\begin{oldquote}}{\end{oldquote}\end{shaded*}}
\usepackage{authblk}
\author[1]{Sagnik Chatterjee}
\author[2]{Vyacheslav Kungurtsev}
\affil[1]{Indraprastha Institute of Information Technology, Delhi (IIITD)}
\affil[2]{Czech Technical University, Prague}
{
    \makeatletter
    \renewcommand\AB@affilsepx{: \protect\Affilfont}
    \makeatother

    \affil[ ]{Email ids}

    \makeatletter
    \renewcommand\AB@affilsepx{, \protect\Affilfont}
    \makeatother

    \affil[1]{sagnikc@iiitd.ac.in}
    \affil[2]{kunguvya@fel.cvut.cz}
}
\date{}
\title{Quantum Solutions to the Privacy vs. Utility Tradeoff}
\begin{document}
\maketitle
\begin{abstract}
In this work, we propose a novel architecture (and several variants thereof) based on quantum cryptographic primitives with provable privacy and security guarantees regarding membership inference attacks on generative models. Our architecture can be used on top of any existing classical or quantum generative models. We argue that the use of quantum gates associated with unitary operators provide inherent advantages compared to standard Differential Privacy based techniques for establishing guaranteed security from all polynomial-time adversaries.
\end{abstract}
\section{Introduction}
The privacy versus accuracy tradeoff in machine learning models has been a central challenge for algorithmic development and understanding. While large-scale generative models such as DALL-E 2~\citep{dalle2022}, Imagen~\citep{imagen2022}, and Stable Diffusion~\citep{stablediff22} have advanced state-of-the-art in terms of the quality of synthetic data beyond previous expectations; their application presents a significant security risk in terms of privacy violation.  The leading methodologies for generative modeling include generative adversarial networks (GANs)~\citep{goodfellow2020generative}, variational autoencoders (VAEs)~\citep{Kingma2013AutoEncodingVB,Rezende2014StochasticBA}, and diffusion models~\citep{ho2020denoising,sohl2015deep,song2020score}. Whereas their intention is to generate data that samples from the population distribution, in practice, all of
these generative model techniques have been shown to “memorize”~\citep{Zhang2021UnderstandingDL} and regenerate the training data ~\citep{carlini2022membership,carlini2023extracting,somepalli2023diffusion,somepalli2023understanding}, which makes them vulnerable to various adversarial attacks such \textit{membership-inference attacks} (MIA)~\citep{Bernau2022AssessingDP,duan2023diffusion,Hilprecht2019MonteCA,hu2023membership,matsumoto2023membership,shokri2017membership,song2021systematic} where an adversary can infer if a given sample was used for training. Such attacks can lead to colossal privacy breaches~\citep{Bommasani2021OnTO}, which must be addressed for safe and reliable operation.

To tackle membership-inference attacks and, more generally, to ensure the privacy of training data, various sophisticated \textit{classical} techniques~\citep{dockhorn2022differentially,matsumoto2023membership,somepalli2023diffusion,lyu2023differentially} have been proposed with differential privacy (DP) based guarantees for diffusion models. However, it has been shown that there is a class of readily implementable adversarial attacks that can recreate training data samples for models satisfying DP-based guarantees \citep{carlini2022membership,ChoquetteChoo2020LabelOnlyMI,Liu2022MembershipIA,song2021systematic}. In many domains of interest, if even a small fraction of the training dataset can be learned by an adversary, then the model cannot be considered private. 
\\\\
\noindent\textbf{Key Idea:} While there are many types of adversarial attacks one can consider, we focus on providing privacy and security guarantees against non-malicious adversarial (NMA) attacks, a broad class that includes MIA. It can be observed from \citep{Bernau2022AssessingDP,carlini2022membership,ChoquetteChoo2020LabelOnlyMI,duan2023diffusion,Hilprecht2019MonteCA,hu2023membership,Liu2022MembershipIA,matsumoto2023membership,shokri2017membership,song2021systematic} that realistic NMA attacks can be modeled by providing adversarial access to the discriminator. Ensuring proper defense against such attacks can be formulated in terms of an interactive game~\cite{carlini2022membership} between 
a challenger and an adversary where the adversary should not be able to determine whether a particular example belongs to the training set even when it is given access to the learning model and the distribution over the data\footnote{MIA can also be viewed through the lense of a hypothesis test~\cite{carlini2022membership}.}.

A novel reinterpretation to this work is that this game can be considered a cryptographic protocol, where we treat the training data as our message and the generated data as a cipher. The goal would be to ensure that the adversary should not be able to glean any meaningful information regarding the training data (our message) given the generated data (our cipher) and black-box access to the training model. This interpretation and its associated analysis uses tools of \textit{security guarantees} such as CPA-security or CCA-security. These, in turn, can be considered to both subsume the weaker, more standard notions of privacy in ML, and provide essential guarantees qualitatively more significant than simply ensuring the privacy of the training data\footnote{Privacy alone does not imply security. For example, one-time pads (OTP) are perfectly private but not secure.}.  
\\\\
\noindent\textbf{Our contribution:} Despite encryption being an extremely natural way to ensure security and privacy, classication encryption techniques are not particularly amenable to obtaining guaranteed accuracy or convergence of training. Therefore, incorporation of classical encryption techniques weights heavily towards complete privacy and no accuracy in the standard privacy-accuracy tradeoff balance. 

In this work, we describe a \textit{quantum} framework for tackling membership-inference attacks based on cryptographic primitives. Working in the quantum regime presents a number of unique computational advantages. Beyond the security advantages which we detail below, under standard complexity-theoretic assumptions, there may exist \textit{quantum-only} distributions that classical generative models may not be able to generate~\citep{chakrabarti2019quantum}. Another motivation to use quantum generative models is the increase in stability and a significant reduction in the number of parameters over their classical counterparts ~\citep{niu2022entangling,stein2020qugan}. 

We also remark that our framework is essentially a generalization of  classical diffusion models. In the quantum framework, our forward process does not rely on the same asymptotic guarantees which classical diffusion models~\citep{ho2020denoising,sohl2015deep,song2020score} rely on, thereby hinting at possible sampling speedups. A classical solution~\citep{xiaotackling} in which non-Gaussian multimodal distributions were used to model the denoising step, could fall prey to the instability issues described in~\citep{arjovsky2017towards,arjovsky2017wasserstein} since there are absolutely no theoretical guarantees regarding the intermediate distributions. In our case, the discriminator needs to distinguish between two quantum states which have identical support. Therefore, the usage of a wide variety of divergences and distances in the discriminator is mathematically advantageous.
\begin{figure}
    \centering
    \includegraphics[width=0.7\textwidth]{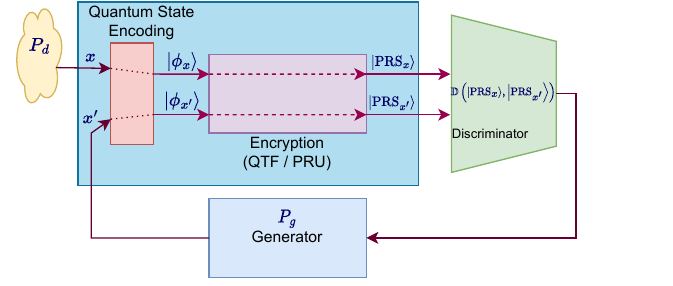}
    \caption{The tuple $(x\sim P_d,x^\prime\sim P_g)$ is first encoded into quantum states and then encrypted. The Discriminator and Generator minimize their losses using the encrypted pair of states.}
    \label{fig:enter-label}
\end{figure}
\section{Background}
An $n$-qubit quantum state is a unit vector on a $2^n$-dimensional complex Hilbert space. The uniform distribution of random quantum states on the $2^n$-dimensional complex Hilbert space is referred to as the \textit{Haar measure over quantum states}. Note that the Haar measure over quantum states is a continuous distribution even when the underlying support is finite. A simple example of a Haar random state is the ensemble 
$
\ket{\psi}_r=2^{-n/2}\sum_{y\in\mathbb{F}_2^n}\left(-1\right)^{r(y)}\ket{y}
$
where $r$ is a truly random function. Since there are $2^{2^n}$ possible choices for $r$, $\ket{\psi}_r$ cannot be generated by polynomial-sized circuits. 

Let $\mathcal{K}=\{\mathcal{K}_n\}_{n\in\mathbb{N}}$ be a sequence of keys from which samples can be efficiently drawn. A quantum-secure pseudorandom function (QPRF) is a family of efficiently computable keyed functions $\mathrm{QPRF}=\{\mathrm{QPRF_k}\}_{k\in\mathbb{N}}:\mathcal{K}_k\times\{0,1\}^n\xrightarrow{}\{0,1\}^n$~\cite{Zhandry2012HowTC} which can be used for the construction of Pseudorandom states. A Pseudorandom quantum state (PRS)~\cite{Ji2018PseudorandomQS,Brakerski2019PseudoRQ} is a quantum state which is information-theoretically indistinguishable from a Haar-random state. In PRS, we replace the truly random function $r$ with an efficiently constructible QPRF $\{f_k\}$ indexed by secret key $k$, such that given query access to $r$ and $f_k:\{0,1\}^n\to \{0,1\}^n$, no efficient polynomially bounded non-uniform quantum adversary $A_Q$ can distinguish between a PRS and a Haar random quantum state.
Therefore, any scheme which involves the construction of a PRS is by definition IND-CPA secure~\cite{Coladangelo2023QuantumTF,Grilo2023EncryptionWQ}. The existence of QPRF assuming one-way functions exist was proved in \cite{Zhandry2012HowTC}. One straightforward candidate for a QPRF function is the family of pseudorandom random permutations (QPRP)~\cite{Ji2018PseudorandomQS}.

A Pseudorandom Unitary (PRU)~\cite{Ji2018PseudorandomQS,Bouland2019ComputationalPT,brandao2016local,Haug2023PseudorandomUA} is a family of unitary operators which is efficiently constructible and indistinguishable from the Haar measure on the unitary group. Quantum Trapdoor functions (QTFs)~\cite{Coladangelo2023QuantumTF} are a class of efficiently constructible unitaries associated with different secret keys. One can use them to form PRS from a classical message $m$ and a secret key $k$. We show how to interpret the QTF construction as a phase-encoding+PRS construction in \cref{sec:tech}. Even though QTFs are unitaries, they are hard to invert without knowing the trapdoor $k$ of the underlying QPRP. 

\section{Technical Overview}\label{sec:tech}
In this section, we outline three constructions that are provably secure against NMA attacks. We denote the real data distribution as $P_d$ and the generated data distribution $P_g$. The crux of the idea is as follows:  
    \begin{center}\label{prop1}
        \emph{Given $x\sim P_d$ and $x^{\prime}\sim P_g$ and a discrimination function $\mathbb{D}$, we want to construct an IND-CPA (or IND-CCA) secure cryptographic scheme with an encryption function $g$ s.t. $$\mathbb{D}\left(x,x^{\prime}\right)=\mathbb{D}\left(g(x),g(x^{\prime})\right)$$}
    \end{center}
For IND-CPA (and IND-CCA) secure cryptographic schemes, the encrypted tuple $g(x),g(x^{\prime})$ must appear pseudorandom w.r.t. to the input tuple $x,x^{\prime}$. It is unclear how to construct classical functions which simultaneously perform IND-CPA secure encryption but also preserve the the discriminative labels in the encrypted tuple. 

A straightforward answer lies in the realm of quantum operators and states. If there exist IND-CPA (or IND-CCA) secure cryptographic schemes with a  unitary(quantum) encryption operator, then by the \textit{distance-preserving property of unitary operators}, all notions of quantum state discrimination are preserved. Formally stated, if $U_g$ is the unitary corresponding to the encryption $g$, and $\mathbb{D}$ denotes the Fidelity measure, then for all pairs of quantum states $\ket{x}$ and $\ket{x^\prime}$, we have 
\begin{equation*}
    \mathbb{D}\left(\ket{g(x)},\ket{g(x^{\prime})}\right)=\mathbb{D}\left(U_g\ket{x},U_g\ket{x^{\prime}}\right)=\bra{x}U_g^\dagger U_g\ket{x^\prime}=\bra{x}\ket{x^\prime}=\mathbb{D}\left(x,x^{\prime}\right)
\end{equation*}
The above argument holds for all distinguishability measures on quantum states as well. 
\begin{remark}
    Even though $\mathbb{D}\left(\ket{g(x)},\ket{g(x^{\prime})}\right)=\mathbb{D}\left(x,x^{\prime}\right)$, the discriminator only has access to the encrypted tuple of states $\left(\ket{g(x)},\ket{g(x^{\prime})}\right)$. Therefore, using the rules of CPA security, there is no polynomially bounded adversary (classical or quantum) who can perform MIA even with non-malicious access to the Generator.
\end{remark}
We now investigate three different constructions that satisfy \cref{prop1}.
\\\\
\noindent\textbf{Phase Encoded PRS Construction}. First, we recall the QTF construction~\cite{Coladangelo2023QuantumTF} briefly.
\begin{defn}[Quantum Trapdoor Function\cite{Coladangelo2023QuantumTF}]
    A QTF is defined as the tuple (GenTR, GenEV, Eval, Invert): $\mathrm{GenTR(1^n)}\xrightarrow{}\mathrm{tr}$. $\mathrm{GenEV(tr)}\xrightarrow{}\ket{eval}=\ket{PRS(\mathrm{tr})}$. $\mathrm{Eval}(\ket{eval},x)\xrightarrow{}\ket{\phi}=Z^x\ket{eval}$. $\mathrm{Invert}(\mathrm{tr},\ket{\phi})\xrightarrow{}x$.
\end{defn}
The classical trapdoor $tr$ is used to construct a quantum public key $\ket{eval}$ which is a PRS. $\ket{eval}$ can now be used to encode a classical string $x$ using a $Z$-twirl operator: $Z^x:=\otimes_{i=1}^n Z^{x_i}$. By properties of QPRF, the state $Z^x\ket{PRS(k)}$ is also a PRS. We can now give the first construction.
\begin{constr}\label{cons:PEPRS}
Given a pair of inputs $x\sim P_d$ and $x^{\prime}\sim P_g$,
    \begin{enumerate}
        \item  Create phase encoded quantum states $\ket{\phi_x}=Z^x\ket{+}^{\otimes n}$ and $\ket{\phi_{x^{\prime}}}=Z^{x^{\prime}}\ket{+}^{\otimes n}$.
        \item Pick a classical trapdoor key $k$ and generate unitary access to a QPRF $\{f_k\}$.
        \item Use the phase kickback trick on $U_{f_k}$ once with with $\ket{\phi_x}$ and once with $\ket{\phi_{x^{\prime}}}$ to obtain the pair $\ket{PRS(x,k)}$ and $\ket{PRS(x^{\prime},k)}$.
        \item Train the Discriminator and Generator on the PRS tuple.
    \end{enumerate}
\end{constr}
Firstly we note that $\ket{PRS(x,k)}$ and $\ket{PRS(x^{\prime},k)}$ are actually pseudorandom states. The proof follows directly from QTF construction~\cite{Coladangelo2023QuantumTF} and the fact that pseudorandom states are invariant under unitary operation. Secondly, we highlight the fact that, unlike any classical crytographic map, $\mathbb{D}\left(\ket{PRS(x,k)},\ket{PRS(x^\prime,k)}\right)=\mathbb{D}\left(\ket{x},\ket{x^\prime}\right)$. Therefore \cref{cons:PEPRS} allows us to train the discriminator properly and securely against adversaries wanting to perform MIA.
\\\\
\noindent\textbf{Parameterized Phase Encoded PRS Construction}. One semantic drawback of \cref{cons:PEPRS} is the fact that we encode the phases \textit{uniformly} across all features (bits of the string in this case). In practice, we might choose to create quantum state encodings of classical feature vectors based on some principles of optimal coding. To incorporate this option, we make use of the Parameterized Z-twirl operator: $RZ(\theta)^x:=\otimes_{i=1}^n RZ(\theta_i)^{x_i}$. Since $RZ(\theta)^x$ is also a diagonal unitary, it commutes with the QTF construction as well.
\begin{constr}\label{cons:PPEPRS}
Given a pair of inputs $x\sim P_d$ and $x^{\prime}\sim P_g$, and a set of feature weights $\theta$,
    \begin{enumerate}
        \item  Create parameterized phase encoded quantum states $\ket{\phi_{x,\theta}}={RZ(\theta)}^x\ket{+}^{\otimes n}$ and $\ket{\phi_{x^{\prime},\theta}}={RZ(\theta)}^{x^{\prime}}\ket{+}^{\otimes n}$.
        \item Generate unitary access to a QPRF $\{f_k\}$ and pick a classical trapdoor key $k$.
        \item Use the phase kickback trick on $U_{f_k}$ once with with $\ket{\phi_x}$ and once with $\ket{\phi_{x^{\prime}}}$ to obtain the pair $\ket{PRS(x,k,\theta)}$ and $\ket{PRS(x^{\prime},k,\theta)}$.
        \item Train the Discriminator and Generator on the PRS tuple.
    \end{enumerate}
\end{constr}
\noindent\textbf{Basis Encoded PRS Construction}. Instead of strictly working with phase-encodings, we may also be interested in basis-encoded quantum states. In order to construct PRS from basis-encoded states, we have to use a construction similar to \cite{Bouland2019ComputationalPT}. We assume oracle access to a PRU $U$, a secret key $k=k_1 k_2\ldots k_T$, where each $k_i\in[4]$, and access to $n$-qubit Pauli operators $\mathcal{O}=\{I_n,X_n,Y_n,Z_n\}$. Given any initial state $\ket{\phi_0}$, the following construction was proven to yield a PRS, provided that the adversary only has black-box access to the PRU $U$.
\begin{equation}\label{basiseqn}
    \ket{PRS(x,k,T)}=U\mathcal{O}_{k_T}U\mathcal{O}_{k_{T-1}}\ldots\mathcal{O}_{k_1}U\ket{\phi_x}
\end{equation}
\begin{constr}\label{cons:BEPRS}
Given a pair of inputs $x\sim P_d$ and $x^{\prime}\sim P_g$, and a parameter $T=\poly{n}$,
    \begin{enumerate}
        \item  Create basis encoded quantum states $\ket{\phi_{x}}$ and $\ket{\phi_{x^{\prime}}}$.
        \item Generate a classical trapdoor key $k$, and obtain unitary access to a PRU $U$.
        \item Construct the PRS tuple $(\ket{PRS(x,k,T)},\ket{PRS(x^{\prime},k,T)})$ as in \cref{basiseqn}.
        \item Train the Discriminator and Generator on the PRS tuple.
    \end{enumerate}
\end{constr}
\section{Discussion}
In this work, we discuss three novel constructions for preventing MIA by leveraging the properties of quantum operators and using privacy and security guarantees from cryptography. Classical analogues of our approach would be difficult to construct, as we discussed. However, since MIA at its heart can be distilled down to a cryptographic game, we believe that there may be Zero-Knowledge-Proof based constructions, or attribute-preserving encryption schemes that could allow our framework to extend to classical techniques. 
\bibliography{refs}
\clearpage
\appendix
\clearpage
\end{document}